# Origin of temperature and field dependence of magnetic skyrmion size in ultrathin nanodots

R. Tomasello[1], K. Y. Guslienko[2,3], M. Ricci[4], A. Giordano[5], J. Barker[6], M. Carpentieri[7], O. Chubykalo-Fesenko[8], G. Finocchio[5]

[1] *Dept. Engineering, Polo Scientifico Didattico di Terni, University of Perugia, 50100 Terni, Italy*

[2] *Depto. Física de Materiales, Universidad del País Vasco, UPV/EHU, 20018 San Sebastián, Spain*

[3] *IKERBASQUE, the Basque Foundation for Science, 48013 Bilbao, Spain*

[4] *Dept. Computer Science, Modeling, Electronics and System Science, University of Calabria, I-87036 Rende, Italy*

[5] *Dept. Mathematical and Computer Sciences, Physical Sciences and Earth Sciences, University of Messina, 98166 Messina, Italy*

[6] *Institute for Materials Research, Tohoku University, 980-8577 Sendai, Japan*

[7] *Dept. Electrical and Information Engineering, Politecnico di Bari, 70125 Bari, Italy*

[8] *Instituto de Ciencia de Materiales de Madrid, CSIC, Cantoblanco, 28049 Madrid, Spain*

PACS:

Understanding the physical properties of magnetic skyrmions is important for fundamental research with the aim to develop new spintronic device paradigms where both logic and memory can be integrated at the same level. Here, we show a universal model based on the micromagnetic formalism that can be used to study skyrmion stability as a function of magnetic field and temperature. We consider ultrathin, circular ferromagnetic magnetic dots. Our results show that magnetic skyrmions with a small radius—compared to the dot radius—are always metastable, while large radius skyrmions form a stable ground state. The change of energy profile determines the weak (strong) size dependence of the metastable (stable) skyrmion as a function of temperature and/or field. These results can open a path toward the design of optimal materials for skyrmion based devices.

Non-linear localized excitations have attracted the attention of physicists for a long time. Such excitations, including solitary waves or solitons, play an important role in optics, quantum field theory, condensed matter and other fields. It is sometimes possible to associate integer numbers (topological charges) to the solitons, which are preserved in their dynamics. Topologically non-trivial magnetization configurations in ferromagnets, such as domain walls, vortices, and skyrmions are currently the focus of a lot of research activity. These spin textures are also candidates for nanoscale device applications—computational paradigms, magnetic storage and programmable logic—due to their small size [1–13].

Skyrmion solutions were obtained first by Skyrme in the non-linear field theory [14]. Subsequently chiral skyrmions were predicted [15], and discovered experimentally in non-centrosymmetric cubic B20 compounds [16–20] which permit an antisymmetric anisotropic interaction, namely the Dzyaloshinskii-Moriya interaction (DMI). This arises from a relativistic correction and relies on spin-orbit interactions [21,22]. Recent efforts have focused on materials with interfacial DMI—especially ultra-thin transition metal/heavy metal multilayers with large spin-orbit coupling such as Co/Pt and Co/Ir [8,23,24]. The DMI, which corresponds to Lifshitz invariants in the micromagnetic energy functional, is necessary to yield axisymmetric skyrmions in ultrathin magnetic elements and the chiral skyrmions can be further stabilized by external magnetic field [8,23,24]. Temperature is usually considered to be detrimental to skyrmion stability, leading to either the transformation of the skyrmion state into a more energetically favorable state [25] or to nucleation of multiple skyrmions and labyrinth domains [8,23,24]. In addition, recent room-temperature experiments with an external out-of-plane magnetic field [8,23], showed a strong non-linear dependence of the skyrmion radius on the external field strength pointing out the key role of the external field.

Rohart *et al.* [26] developed a domain wall model of Néel skyrmions at zero temperature $T$ and zero magnetic field $H_{ext}$, and identified a critical DMI parameter $D_c$ quantifying the influence of DMI that is independent of the dot size. They found that a single skyrmion is metastable for DMI magnitudes $|D|$ smaller than $D_c$ and does not exist if $|D| > D_c = (4/\pi)\sqrt{AK_{eff}}$, where $A$ and $K_{eff}$ are the exchange stiffness and effective uniaxial magnetic anisotropy constants. Within their analytic model, the skyrmion radius $R_{sk}$ diverges as $D \to D_c$. In a finite magnetic dot, however, $R_{sk}$ is determined by the confining potential and cannot exceed the dot radius $R_d$ [26]. This analytical model yields incorrect results for $D > D_c$ and should be improved.

The temperature dependence of the skyrmion radius in an infinite film was calculated numerically by Barker *et al.* [27] who attempted to use Rohart's expression for $R_{sk}$ with the temperature dependence of the magnetic properties of the host material to explain the behavior. In that paper [27], the bulk scaling relations of $A$ and $K_u$ (uniaxial anisotropy constant) with respect to the reduced magnetization $m(T) = M_s(T)/M_s(0)$ were used ($M_s$ is the saturation magnetization), while no temperature dependence of $D$ was considered. The authors concluded that in ferromagnets the skyrmion radius has a weak quasi-linear temperature dependence. This is rather a result of using the Rohart´s *et al.* equation for small radius (~10 nm) skyrmions which have a size comparable with the domain wall width.

Our work is motivated by the lack of understanding the skyrmion size and stability as a function of control parameters such as temperature and external field in a finite-size sample. Here, we provide a simple analytic tool for developing materials and geometries which can host skyrmions in a target temperature and field window. Particularly, the analytical model predicts weak/strong dependence of skyrmion size on control parameters which is further confirmed by numerical modeling. This would provide a simple analytic tool for developing materials and geometries which can host skyrmions in a target temperature and field window.

Here, we develop a theoretical approach to skyrmion stability based on minimization of the magnetic energy using a skyrmion ansatz in confined geometries. We show that the thermal evolution of skyrmion size in ultrathin dots is best expressed in a $Q$-$d$ phase space of the skyrmion quality factor $Q = 2K_u/\mu_0 M_s^2$ ($\mu_0$ is the vacuum permeability), and the reduced DMI strength $d = |D|l_{ex}/A$ ($l_{ex} = \sqrt{2A/\mu_0 M_s^2}$ is the exchange length). The proximity of the point describing the skyrmion state in this $Q$-$d$ phase space, to the border of the skyrmion stability, is the crucial parameter in understanding the linear or non-linear change in radius with temperature or field as observed in micromagnetic simulations. The temperature dependence of the skyrmion stability is expressed as scaling relations of both $Q$ and $d$ with changing $m(T)$. We find the equilibrium skyrmion size increases rapidly with temperature only when approaching a critical line $d_c(Q) = (4/\pi)\sqrt{Q-1}$. The size becomes comparable with the dot radius at $d(T) > d_c(T)$. The metastable skyrmions have small radius ($R_{sk} \ll R_d$), whereas the stable (ground state) ones have large radius. The out-of-plane external magnetic field can be used to bring metastable skyrmions closer to or farther from the critical line $d_c(Q)$ where the metastable skyrmion becomes a ground state skyrmion.

The magnetic skyrmion can be described by the energy functional $E[\mathbf{m}] = \int dV \varepsilon(\mathbf{m})$, with the energy density

$$\varepsilon(\mathbf{m}) = A(\nabla \mathbf{m})^2 + \varepsilon_{DMI} + K_u m_z^2 - 0.5 M_s \mathbf{m} \cdot \mathbf{H}_m - M_s \mathbf{m} \cdot \mathbf{H}_{ext}, \qquad (1)$$

where $\mathbf{m} = \mathbf{M}/M_s$ is the unit magnetization vector ($M_s$ is the saturation magnetization at given *T*), *A* is the exchange stiffness constant, $\varepsilon_{DMI} = D[m_z(\nabla \cdot \mathbf{m}) - (\mathbf{m} \cdot \nabla)m_z]$ is the interface DMI energy density, with *D* being the DMI parameter, $K_u$ is the perpendicular uniaxial anisotropy constant, $m_z$ is the magnetization *z*-component, $\mathbf{H}_m$ is the magnetostatic field, and $\mathbf{H}_{ext}$ is the external magnetic field.

We assume that in an ultrathin circular dot the magnetization $\mathbf{m}$ does not depend on the thickness (*z*-coordinate). We introduce polar coordinates in the dot plane $\boldsymbol{\rho} = (\rho, \phi)$, and define spherical angles of the magnetization vector $(\Theta, \Phi)$ as functions of $\boldsymbol{\rho}$. We also assume that the static skyrmion centered in the dot is axially symmetric and use the equations $\Theta_0(\boldsymbol{\rho}) = \Theta_0(\rho)$, $\Phi_0(\boldsymbol{\rho}) = \phi + \phi_0$. In seeking a general expression for the dependence of the skyrmion size on both temperature and external magnetic field, and to overcome drawbacks of previous approaches in estimating the skyrmion radius [28](see note 1 in the Supplemental Material), we developed an analytical approach using a different skyrmion ansatz from Ref. [26]. To minimize the energy functional (1), we propose the trial function

$$\tan \frac{\Theta_0(r)}{2} = \frac{r_{sk}}{r} e^{\xi(r_{sk} - r)}, \qquad (2)$$

where $r = \rho/l_{ex}$ $\xi^2 = Q - 1$, and the polar angle $\Theta_0$ describes the static skyrmion profile. The accuracy of the ansatz (2) was numerically verified in the case of 2D easy axis infinite ferromagnets (*Q*>1) in Ref. [29]. Eq. (2) recovers the Belavin-Polyakov solution for the isotropic case $\xi = 0$ and leads to the finite exchange energy at $r \approx 0$. Eq. (2) is also very similar (in the area of its applicability, at $r \approx r_{sk} \gg 1$) to the ansatz used in the theory of bubble domains [30] in highly anisotropic ferromagnetic films [28] (see Approach 3 in note 1 in the Supplemental Material).

The energy $E$ and the equilibrium skyrmion radius $r_{sk} = R_{sk} / l_{ex}$ can now be expressed as functions $E[r_{sk}, Q, d, H_{ext}]$ and $r_{sk}[Q, d, H_{ext}]$. The conditions of the skyrmion stability can be found using standard variational procedure; solving the equations $\partial E / \partial r_{sk} = 0$, $\partial^2 E / \partial r_{sk}^2 = 0$. To account for the temperature dependence of the skyrmion radius, we use the scaling approach for macroscopic (micromagnetic) parameters which usually decrease with temperature increasing due to the magnetization fluctuation effects. Including thermal effects into a micromagnetic approach is known to largely overestimate the Curie temperature as compared to the more accurate atomistic approach [31]. However, the largest temperatures considered in the present study are very far from the Curie temperature ($T/T_c < 1/2$). In this case, the micromagnetic approach produces accurate results in terms of the thermodynamically averaged quantities as a function of the magnetization $m(T)$. We calculated the temperature dependence of uniaxial magnetic anisotropy $K_u(T) = K_u(0) m(T)^\gamma$ achieving $\gamma = 3.03$ that is close to the Callen-Callen relation [32]. The scaling of the exchange $A(m) = A(0) m^\alpha$ and DMI $D(T) = D(0) m(T)^\beta$ parameters are found by atomistic simulations of an infinite thin film calculating the thermal spin wave spectrum and fitting the long wavelength regime (small **k**-vectors) with a linear spin wave theory [28] (see note 2 in the Supplemental Material). We obtain $\alpha = \beta = 1.5$, consistent with other recent results [33,34].

To benchmark the theory, we consider a circular nanodot (e.g., Co) of diameter $2R_d = 400$ nm and thickness of 0.8 nm assumed to be in contact with a thin layer of heavy metal (e.g., Pt) giving an appreciable interfacial DMI. We performed systematic micromagnetic simulations to calculate the skyrmion size as a function of the out-of-plane external field $H_{ext}$ (from 0 mT to 50 mT) and temperature (from 0 K to 300 K) integrating the Landau-Lifshitz-Gilbert equation of motion for the reduced magnetization $\mathbf{m} = \mathbf{M}/M_s$ [28] (see note 3 in the Supplemental Material). At $T$=0 K, we used the following material parameters: $M_s$=600 kA/m [8], $A$=20 pJ/m [35], $D$=3.0 mJ/m$^2$ [36,37], $K_u$=0.60 MJ/m$^3$ [8], and Gilbert damping $\alpha_G$=0.1 [38]. As a reference, Rohart's critical DMI value $D_C$=3.48 mJ/m$^2$ for our parameters at $T$=0 K [26].

Temperature causes the skyrmion to diffuse and leads to fluctuations of the diameter and deformations of the skyrmion shape [27,39,40], losing the circular symmetry. We therefore calculated the effective diameter by assuming that the area of the skyrmion core (here it is the region where the *z*-component of the magnetization is negative) is equivalent to a circle [27]. The skyrmion diameter and perimeter display approximately Gaussian statistical distributions [28] (see note 3 in the Supplemental Material) with increasing widths with temperature. The application of the magnetic field significantly decreases both mean and standard deviation of the distributions.

Fig. 1(a) compares the magnetic field dependence of the skyrmion diameter calculated by micromagnetic simulations with the analytic skyrmion ansatz, Eq. (2), at $T=0$ K. There is a reasonably good agreement considering there are no fitting parameters (maximum difference around 8% at zero field). Fig. 1(b) shows the temperature dependence of the skyrmion diameter calculated with micromagnetic simulations for three values of the external field (open symbols). At zero field, the skyrmion diameter rapidly expands with increasing temperature. The increasing out-of-plane field causes a weak quasi-linear increasing of the diameter up to $T=300$ K (red circles and green triangles in Fig. 1(b)). Our computations show that two thermal/field regimes exist: at high temperature ($T>200$ K) and low field ($H_{ext}<10$ mT), the skyrmion size is strongly influenced by thermal fluctuations, whereas at low temperature ($T\leq200$ K) independent of the field, or at high temperature and high field ($H_{ext}\geq10$ mT), the skyrmion is weakly affected by thermal fluctuations. These different behaviors were not observed in Ref. [27], because with the parameters used, the skyrmions were always in the metastable region. Calculating the skyrmion diameter as a function of the external field for $T=300$K (Fig. 1(c)) there is a large, non-linear expansion of the diameter and larger variance as the external field is reduced. This is consistent with the experimental results in Refs. [8,23].

By including the scaling relations (calculated from atomistic simulations) in the analytical approach, the results are in agreement with micromagnetic simulations until the skyrmion is in the metastable region far from the boundary of the stable region [28](e.g. when $T<200$ K for $H_{ext}=0$ mT, see note 3 in the Supplemental Material). When approaching the ground state region, the confining potential—due to the magnetostatic field and the DMI boundary conditions—starts to play a significant role in fixing the skyrmion size [11,26], and the skyrmion diameter depends on the dot size. To account for this, we consider the scaling exponent $\gamma$ of the uniaxial perpendicular anisotropy $K_u(T) = K_u(0) m(T)^\gamma$ as a fitting parameter. This is because the analytical model is developed within the thin-film approximation where the effective anisotropy is computed as $K_{eff}(T) = K_u(T) - 0.5\mu_0 M_s(T)^2$, but the numerical micromagnetic solution includes the full magnetostatic calculation. By performing athermal (deterministic) micromagnetic simulations [28] (see note 3 in Supplemental Material), an excellent agreement is found for $\gamma=3.585$. With this value, we can calculate the skyrmion diameter dependence on temperature and field by Eq. (2) (dashed lines in Fig. 1(b)). The analytical (ansatz Eq. (2)) and micromagnetic results agree well and we conclude that our analytical expression is accurate in predicting the skyrmion size for arbitrary temperature and external field combinations.

Fig. 1(c) shows how even a small external field strength $H_{ext}$=5 mT significantly reduces the skyrmion size at room temperature. This occurs because the magnetization region outside of the skyrmion core expands—the field direction is opposite to the skyrmion core magnetization—leading to a shrinking of the skyrmion. The non-linear field-skyrmion size dependence is in qualitative agreement with recent experimental results (see, for comparison, Fig. 4a in Ref. [23] and Fig. S8 in Supplemental Material of Ref. [8]).

To understand the origin of the two thermal/field regimes, we focus on the data obtained at zero external field. We calculate the critical DMI parameter for each set of the scaled parameters by the expression $D_c(T) = 4\sqrt{A(T) \cdot K_{eff}(T)}/\pi$ [26]. When $D$ approaches $D_c$ (as temperature increases, see Fig. 2), a sharp increase of the skyrmion size $R_{sk}(T)$ occurs [26]. This could explain why, at room temperature, the skyrmion size increases for $H_{ext}$<5 mT exhibiting a non-linear dependence on the external field [8,23] (see Fig. 1(c)).

The thermal scaling of the macroscopic parameters leads to the temperature dependence of the quality factor $Q(T) = Q(0)m(T)$. Taking into account the definition $d(T) = D(T)l_{ex}(T)/A(T)$, we derive the temperature dependence of the reduced DMI parameter as $d(T) = d(0)[m(T)]^{\beta-\alpha/2-1}$. Note that temperature dependence of the reduced critical DMI parameter $d_c(m) \propto m^{-1}\sqrt{K_{eff}(m)} \propto \sqrt{Q(m)-1}$ includes neither the exchange stiffness $\alpha$ nor the DMI exponent $\beta$. This justifies the use of the skyrmion stability diagram expressed in the reduced coordinates, $Q$ and $d$—presented in Fig. 3(a). The parameters at $T$=0 K are: $Q(0)$=2.65, $d(0)$=1.41, $l_{ex}(0)$=9.4 nm, (the scaling law for the reduced DMI parameter is $d(m) \propto m^{\beta-\alpha/2-1}$, i.e., $d(m) \propto m^{-0.84}$). The dependence $d(m)$ determines the evolution of the point describing the skyrmion configuration in the $Q$-$d$ plane accounting for the change of temperature via the $m(T)$ law (Fig. 3(a)). The increase of parameter $d$ leads to the stabilization of the skyrmion state and a rapid expansion of the skyrmion radius. This effect is especially strong at $H_{ext}$=0 mT. A finite value of the field $H_{ext}$ opposing the skyrmion core magnetization, results in a contraction of the skyrmion radius and the skyrmion radius weakly increases with $T$. In other words, the skyrmion radius increase is suppressed by the finite magnetic field which does not occur at zero field.

There are two qualitatively different behaviors of how the skyrmion radius $R_{sk}(T)$ changes with temperature at $H_{ext}$=0 mT according to the parameters $Q$ and $d$:

1) The skyrmion radius $R_{sk}(T)$ is an almost constant function of temperature when the skyrmion configuration remains in the metastable region.
2) The skyrmion radius $R_{sk}(T)$ increases sharply with increasing temperature when approaching the region of skyrmion ground state stability.

The first scenario is realized when the $T=0$ K parameters $Q(0)$, $d(0)$, $l_{ex}(0)$ correspond to the skyrmion metastable state (Figs. 3(a) and (c)). In this case, the skyrmion radius at $T=0$ K is small and has only a weak temperature dependence $R_{sk}(T)$ up to 200 K because the skyrmion is deep in the region of metastability in $Q$-$d$ phase diagram. The decreasing function $d(m)$ guarantees that the skyrmion state stays in the ($Q$, $d$)-area of the skyrmion metastability for increasing temperature.

The second behavior occurs when a skyrmion, initially in the metastable state with parameters $Q(0)$, $d(0)$, $l_{ex}(0)$ (see Figs. 3(a) and (b)), moves towards the area of global stability in the $Q$-$d$ space. This is realized for our dot magnetic parameters from $T=200$ K to $T=350$ K (green points in Fig. 3 (a)). In this case, the skyrmion radius at $T=0$ K is small, but it shows a strong dependence with increasing temperature $R_{sk}(T)$.

For further increasing of $T$ up to 400 K, the skyrmion reaches the area of its stability (magenta point in Fig. 3(a)) crossing the uniform state-skyrmion equilibrium line (red dashed line in Fig. 3(a)). At this line the skyrmion energy is equal to the perpendicular uniform state energy, $d = d_c$, the skyrmion radius is large and depends on the confining potential. In particular, the skyrmion radius calculated from deterministic micromagnetic simulations with scaled values of the parameters (corresponding to $T=400$ K) is much larger than the skyrmion radius for $T=300$ K (compare spatial distribution of the magnetization in Fig. 3(b) with the one in Fig. 3(c)). In the region of stability where the skyrmion is the ground state of the ferromagnet, the ratio $R_{sk}/R$ is a weak function of all parameters.

On the other hand, when considering in the micromagnetic simulations thermal fluctuations at $T=400$ K, we observe large deformations of the skyrmion which turns into a "horseshoe"-like configuration as already observed in experiments [8], while it remains quasi-circular for $T=300$ K (see insets in Fig. 3(a)). We do not consider here stabilization of multiple skyrmions or labyrinth domain textures, assuming that the parameter $d$ is not very large and we are always in the area of the single skyrmion stability.

Understanding how to move skyrmions between the metastability and the confined ground state can be exploited to enhance the electrical skyrmion detection [40]. Fig. 4 shows a racetrack memory device where the skyrmions in the track are metastable and therefore small—giving a high storage density. The skyrmions are detected under a magnetic tunnel junction (green square) with a polarized layer with a magnetization pointing along the out-of-plane direction which generates a dipolar field parallel to the skyrmion core magnetization. This field can modify the stability properties of the skyrmion in the region below the contact, moving it through the $Q$-$d$ phase space. By shifting the skyrmions across the line of stability, their radius will expand significantly making it much easier to detect from the tunnel magnetoresistance signal (see Supplemental Material Video 1 [28]). After leaving the detection regions, the skyrmions will return to their small size in the metastable region of the $Q$-$d$ diagram.

In summary, we developed a theoretical framework describing the skyrmion stability in ultrathin dots as a function of the external magnetic field and temperature by combining a proper ansatz with thermal scaling relations of the micromagnetic parameters $A$, $K_u$ and $D$. We showed that the strong temperature dependence of the skyrmion diameter occurs because the thermal evolution of an initially metastable skyrmion brings it toward the region where the skyrmion is the ground state due to increase of the effective DMI strength in comparison to the anisotropy. Our results, corroborated by extensive micromagnetic simulations, provide a tool, the $Q$-$d$ phase diagram, to determine the transition from the metastable to ground state skyrmion configuration, as well as the skyrmion size in presence of both out-of-plane external field and temperature. Our achievements can drive the design of racetrack memories where localized manipulation of magnetic parameters can be used to vary the skyrmion size and stability, improving the ability for the skyrmion electrical detection.

**ACKNOWLEDGMENTS**

This work was supported by the program of scientific and technological cooperation between Italy and China (code CN16GR09) funded by Ministero degli Affari Esteri e della Cooperazione Internazionale, and by the bilateral Italy-Turkey project (CNR Grant #: B52I14002910005, TUBITAK Grant # 113F378). R.T. and M.R. thank Fondazione Carit Projects 'Sistemi Phased-Array Ultrasonori', and 'Sensori Spintronici'. K.G. acknowledges support by IKERBASQUE (the Basque Foundation for Science). The work of K.G. and O.C.-F. was supported by the Spanish Ministry of Economy and Competitiveness under projects MAT2013-47078-C2-1-P, MAT2013-47078-C2-2-P and FIS2016-78591-C3-3-R. J.B. acknowledges support from the Graduate Program in Spintronics, Tohoku University.

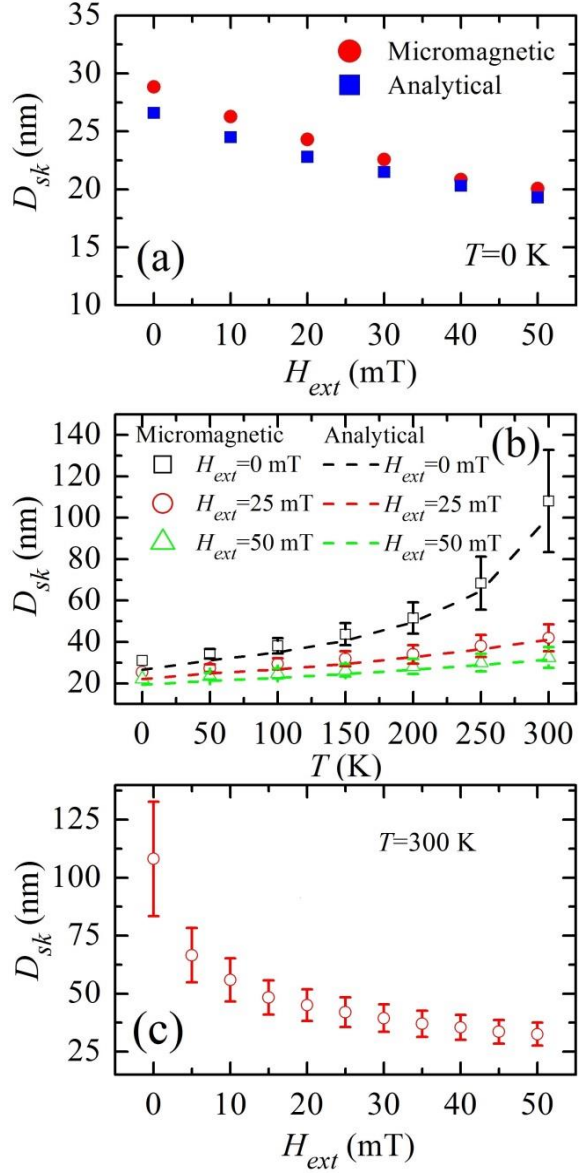

FIG. 1: (a) Skyrmion diameter as a function of the perpendicular external field as computed by micromagnetic simulations (red circles) and by analytical computations based on Eq. (1) (blue squares) at $T=0$ K. (b) Skyrmion diameter as a function of temperature for three values of the perpendicular external field. The symbols represent the mean value of the skyrmion diameter as obtained by micromagnetic simulations including thermal fluctuations, where the error bar corresponds to the standard deviation, while the dashed curves are related to the analytical results with scaled values of the macroscopic parameters, using $\alpha=1.50$, $\beta=1.50$, and $\gamma=3.585$. $\gamma$ is approximatively 15% larger than the one derived from atomistic simulations [32]. (c) Mean value of the skyrmion diameter and corresponding standard deviation as error bar as a function of applied external field for $T=300$ K, calculated from micromagnetic simulations.

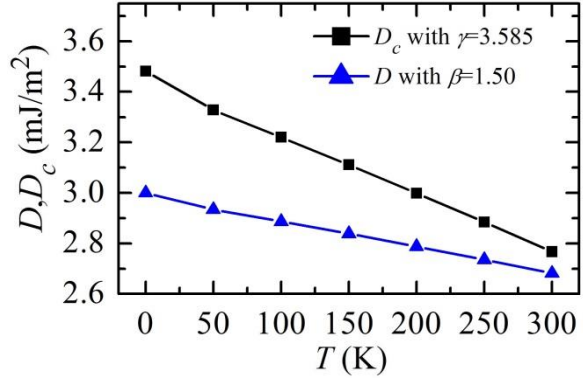

FIG. 2: Comparison of temperature dependences of the scaled DMI parameter $D$ and the scaled critical DMI parameter $D_c$ when $H_{ext}$=0 mT.

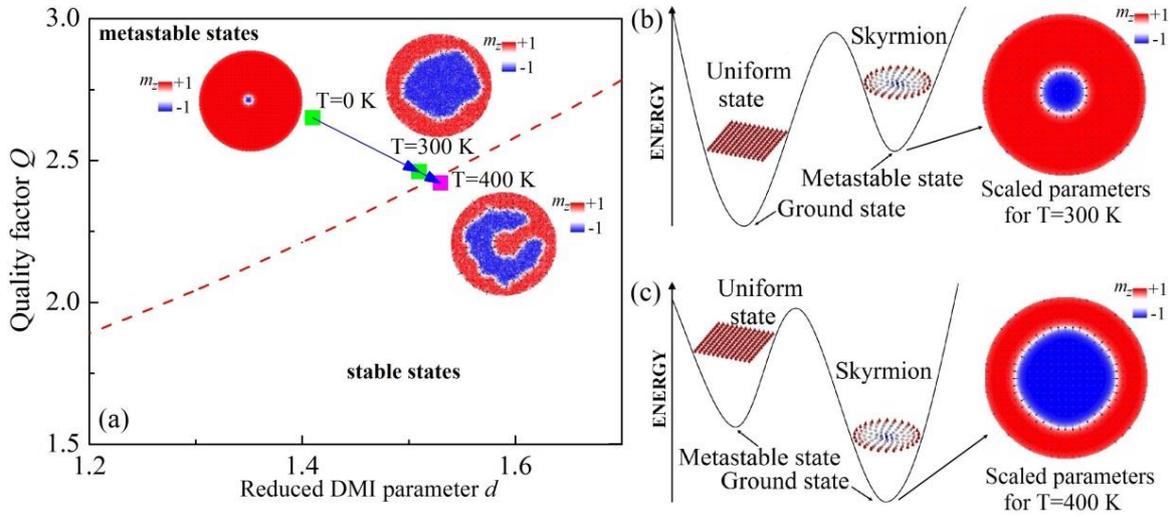

FIG. 3: (a) $Q$-$d$ skyrmion stability diagram for $H_{ext}$=0 mT. The green squares indicate the metastable skyrmion states at $T$=0 K and $T$=300 K, while the magenta square indicates the stable skyrmion state at $T$=400 K obtained with our parameters. The dashed red line represents the critical DMI parameter $D_c$ as defined in Ref. [26]. The blue arrows indicate how the state evolves as a function of temperature. Insets: example of spatial distributions of the magnetization at $T$=0, 300 and 400 K. Sketch of the energy profile of the dot when the skyrmion corresponds to (b) a metastable state and (c) when it occupies an absolute energy minimum (ground state). Two spatial distributions of the magnetization shown in (b) and (c) refer to deterministic full micromagnetic simulations with scaled values of the parameters for $T$=300 K and $T$=400 K, respectively. For all the spatial distributions of the magnetization, a color scale linked to the normalized $z$-component of the magnetization is also illustrated.

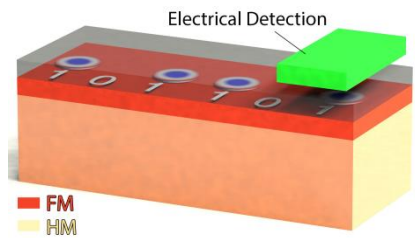

FIG. 4: (a) Sketch of a racetrack memory device where the skyrmions (bit '1') are electrically detected below an MTJ (green square).

# SUPPLEMENTARY INFORMATION

# Origin of temperature and field dependence of magnetic skyrmion size in ultrathin nanodots

R. Tomasello[1], K. Y. Guslienko[2,3], M. Ricci[4], A. Giordano[5], J. Barker[6], M. Carpentieri[7], O. Chubykalo-Fesenko[8], G. Finocchio[5]

[1] Department of Engineering, Polo Scientifico Didattico di Terni, University of Perugia, 50100 Terni, Italy

[2] Depto. Física de Materiales, Universidad del País Vasco, UPV/EHU, 20018 San Sebastian, Spain

[3] IKERBASQUE, the Basque Foundation for Science, 48013 Bilbao, Spain

[4] Department of Computer Science, Modelling, Electronics and System Science, University of Calabria, via P. Bucci, I-87036 Rende (CS), Italy

[5] Department of Mathematical and Computer Sciences, Physical Sciences and Earth Sciences, University of Messina, 98166 Messina, Italy

[6] Institute for Materials Research, Tohoku University, 980-8577 Sendai, Japan

[7] Department of Electrical and Information Engineering, Politecnico di Bari, 70125 Bari, Italy

[8] Instituto de Ciencia de Materiales de Madrid, CSIC, Cantoblanco, 28049 Madrid, Spain

## Supplementary note 1

We assume that in an ultrathin circular dot with thickness *t* about 1 nm and radius $R_d$, the magnetization **m** does not depend on the thickness (*z*-coordinate). We introduce the polar coordinates in the dot plane $\boldsymbol{\rho} = (\rho, \phi)$, and define spherical angles of the magnetization vector $(\Theta, \Phi)$ as functions of $\boldsymbol{\rho}$. We also assume that the magnetization of the static skyrmion centered in the dot is circularly symmetric and use the equations $\Theta_0(\boldsymbol{\rho}) = \Theta_0(\rho)$, $\Phi_0(\boldsymbol{\rho}) = \phi + \phi_0$. The DMI energy depends on the product of *D* and skyrmion chirality $C = sign(m_\rho) = \cos\phi_0$ demanding $\phi_0 = 0, \pi$ (Néel skyrmions) and $CD = -|D| < 0$. The energy of the face magnetic charges is accounted via the effective uniaxial anisotropy $K_{eff} = K_u - \mu_0 M_s^2 / 2$. The magnetostatic energy of the volume and side surface charges is neglected due to small dot thickness.

The polar angle $\Theta_0$, describing the static skyrmion profile, satisfies the equilibrium equation obtained from the minimization of the energy functional:

$$\nabla^2 \Theta_0 = \left(\xi^2 + \frac{1}{r^2}\right)\sin\Theta_0 \cos\Theta_0 - \frac{d}{r}\sin^2\Theta_0 + h\sin\Theta_0, \qquad (S1)$$

where $\xi^2 = Q - 1 = 2K_u / \mu_0 M_s^2 - 1$ describes the effective easy-axis magnetic anisotropy, $h = H_{ext}/M_s$ is reduced perpendicular magnetic field, $r = \rho/l_{ex}$, and $d = |D|l_{ex}/A$ ($l_{ex} = \sqrt{2A/\mu_0 M_s^2}$ is the exchange length). The non-linear differential Eq. (S1) should be complemented by the boundary conditions $\Theta_0(0) = \pi$ or 0 (i.e., $m_z = -1$ or $+1$), $\Theta_0(r_{sk}) = \pi/2$ and $\partial\Theta_0/\partial r = d/2$ at $r = R_d/l_{ex}$ [1]. Here $r_{sk} = R_{sk}/l_{ex}$ is the reduced skyrmion radius defined as $m_z(r_{sk}) = 0$.

In order to obtain the skyrmion radius dependence on field and temperature, we can consider three state-of-the-art approaches. The first one solves Eq. (S1) in some limiting case, while the other two minimize the energy functional (Eq. (1) in the main text) using different trial functions.

*Approach 1.*

Eq. (S1) has a simple analytical solution only for the dominating isotropic exchange interaction ($Q=1$, $d=0$), $\tan(\Theta_0(r)/2) = r_{sk}/r$, which is known as Belavin-Polyakov (BP) soliton.

We use the definite skyrmion polarization $p=-1$ defined as the $z$-component of the magnetization in the skyrmion center, $p = \cos\Theta_0(0)$.

*Approach 2.*

The simplest trial function widely used in the domain wall theory is the linear ansatz ( $\Theta_0(\rho) = \pi(1-\rho/2R_{sk})$ at $\rho \leq R_{sk}$, $\Theta_0(\rho) = 0$ at $\rho \geq 2R_{sk}$ [2]. The equilibrium skyrmion radius derived from the linear ansatz is directly proportional to the DMI constant

$$R_{sk}(H_{ext}) = \frac{D}{4M_s^2\left[Q - 1 + \frac{1}{\pi}\left(1 - \frac{4}{\pi^2}\right)\frac{H_{ext}}{M_s}\right]}, \tag{S2}$$

where the perpendicular field $H_{ext}$ is opposite to the skyrmion core magnetization.

It can be shown that the skyrmion radius given by Eq. (S2) can describe neither the field dependence obtained by micromagnetic simulations and experimental measurements at room temperature, nor the simulated temperature dependence of the skyrmion radius at zero field. The field and temperature dependence of the skyrmion radius given by Eq. (S2) exhibits a different slope in comparison with obtained by micromagnetic simulations (see Fig. S1) and measured experimentally. In addition, the drawbacks of the linear ansatz are: 1) it predicts that the chiral skyrmion is metastable at any $|D| > 0$ (the critical value of DMI $D_c = 0$) and any $Q > 1$, the skyrmion cannot be the dot ground state; 2) the area of the skyrmion metastability ($|D| > 0$, $Q > 1$) does not depend on the exchange constant $A$ for small radius skyrmions, $R_{sk}/R \leq 1/2$; 3) the dependence of the skyrmion radius $R_c(D)$ calculated within the ansatz is linear in $D$, whereas this function is strongly non-linear with an inflation point at $D \approx D_c$ (with a finite $D_c$) as also shown numerically in Ref. [1].

*Approach 3.*

We can choose the following ansatz describing a circular domain wall:

$$\tan\frac{\Theta_0(\rho)}{2} = e^{p\frac{(\rho - R_{sk})}{\Delta}} \tag{S3}$$

where $\Delta = \sqrt{A/K_{eff}}$ is the domain wall width.

Eq. (S3) is a very good approximation within the limit $R_{sk} \gg \Delta$. However, in our case, $R_{sk}/\Delta \approx 2$ only, and then Eq. (S3) has a bad accuracy and cannot be applied. Moreover, the ansatz (S3) leads to a singularity in the exchange energy at $\rho \to 0$, because it does not satisfy the boundary condition $\Theta_0(0) = \pi$. Substituting the ansatz (S3) to the magnetic energy functional (Eq. (1) in the main text), and conducting integration in the vicinity of $\rho \approx R_{sk}$, the equilibrium skyrmion radius of Ref. [1] is obtained

$$R_{sk} = \frac{\Delta}{\sqrt{2(1 - D/D_c)}}, \qquad (S4)$$

Eq. (S4) is applicable in a narrow region at $D < D_c$, when the skyrmion is metastable/unstable. Other drawback of Eq. (S4) is a prediction of such a small value of $R_{sk} \approx \Delta$, for which the ansatz (S3) cannot be applied.

Therefore, the form of the approximate solution for the skyrmion equilibrium radius $R_{sk}$ depends on the relative value of the ratio $\varsigma = R_{sk}/\Delta$:

1) The Belavin-Polyakov model is approximately valid at $\varsigma \leq 1$ (small radius skyrmion);
2) The linear $\Theta_0(\rho)$ model and Eq. (S2) are valid for intermediate $1 < \varsigma \leq 5$ (however, it has other drawbacks);
3) The domain wall ansatz (S3) is valid for large $R_{sk}$, $\varsigma > 5$.

## Supplementary note 2

The temperature dependence of macroscopic magnetic parameters, such as $m(T)$ and $K_u(T)$, can be calculated from a microscopic formalism. This formalism includes spin fluctuations which cause the macroscopic magnetization to change length and the magnetic parameters to decrease as the system samples larger areas of the free energy [3].

We used atomistic spin dynamics to calculate the temperature dependence of $A(T)$, $K_u(T)$ and $D(T)$ from the finite temperature spin wave spectrum. Temperature is included in the classical (white noise) limit and general details of the method can be found in Ref. [4]. We consider a 2D simple cubic thin film in the *x-y* plane, as was used in the micromagnetic calculations. The Hamiltonian is:

$$H = -J\sum_{<ij>} \mathbf{S_i} \cdot \mathbf{S_j} + \sum_{<ij>} d\left(u_{ij} \times \hat{z}\right) \cdot \left(\mathbf{S_i} \times \mathbf{S_j}\right) - \mu_S \sum_i \mathbf{B} \cdot \mathbf{S_i} \quad , \tag{S5}$$

where $\mathbf{S_i}$ are classical spins of unit length, $J = 1.0 \cdot 10^{-20}$ and $d = 6.0 \cdot 10^{-22}$ Joules, and $\mu_S = 1.62$ $\mu_B$ corresponding to the micromagnetic parameters with the unit cell sizes *a*=0.25 nm and *c*=0.4 nm. The angle brackets *<ij>* indicate a summation over nearest neighbors only. We use a large magnetic field $H_{ext}$=(0, 10T, 0) to force the magnetization to lie in the plane of the film while also maintaining a uniform texture. The DMI manifests as a k-shift of the spin wave frequency. The system size is 256x256x1 unit cells, i.e., we conduct simulations in the 2D limit and with a large in-plane size to reduce finite size effects. Periodic boundary conditions are used in the *x-y* directions.

The dynamics are solved using the Landau-Lifshitz-Gilbert equation for each spin using the Heun method with the Gilbert damping $\alpha_G$=0.001 and $\Delta t$=1 fs.

The finite temperature spin wave spectrum is calculated from the space-time Fourier transform of the $S_x$, $S_y$ spin component fluctuations (following the same procedure as in Ref. [5]) giving the result in Fig. S2.

To calculate the change in exchange stiffness and DMI, we fit the spin excitation spectrum with the linear spin wave dispersion relation (see Fig. S3)

$$\omega(\mathbf{k}) = \gamma_0 \left( H + \nu(T)\frac{4J}{m}(1 - g_k) + \zeta(T)\frac{2d}{m}k_x g_k \right) \tag{S6}$$

where $m(T) = M_s(T)/M_s(0)$ is the reduced magnetization ($M_s$ is the saturation magnetization), $g_k = (\cos ak_x + \cos ak_y)/2$ for the 2D simple cubic lattice. $\nu(T)$ and $\zeta(T)$ describe the effective temperature dependence $A(T)/A(0)$, and $D(T)/D(0)$, respectively.

The scaling laws are then found by fitting the relations (see Fig. S4)

$$\frac{A(T)}{A(0)} = \left(\frac{M(T)}{M(0)}\right)^\alpha \qquad \frac{D(T)}{D(0)} = \left(\frac{M(T)}{M(0)}\right)^\beta \tag{S7}$$

giving the exponents $\alpha$=1.498 +/- 0.001 and DMI $\beta$=1.495 +/- 0.007. These results are consistent with literature [6,7]. In particular, they depend on the lattice structure and generally are material-specific. Since the micromagnetic simulations are conducted on a square lattice, the bulk scaling exponent $\alpha$ is expected to be approximatively 1.6 [6].

## Supplementary note 3

The micromagnetic computations are carried out by means of a state-of-the-art micromagnetic solver which numerically integrates the Landau-Lifshitz-Gilbert (LLG) equation by applying the time solver scheme Adams-Bashforth [8]:

$$(1+\alpha_G^2)\frac{d\mathbf{m}}{d\tau} = -(\mathbf{m}\times\mathbf{h}_{eff}) - \alpha_G \mathbf{m}\times(\mathbf{m}\times\mathbf{h}_{eff}), \quad (S8)$$

where $\alpha_G$ is the Gilbert damping, $\mathbf{m} = \mathbf{M}/M_s$ is the normalized (reduced) magnetization, and $\tau = \gamma_0 M_s t$ is the dimensionless time, with $\gamma_0$ being the gyromagnetic ratio, and $M_s$ the saturation magnetization. $\mathbf{h}_{eff}$ is the normalized effective magnetic field, which includes the exchange, magnetostatic, anisotropy and external fields, as well as the interfacial DMI and the thermal field. The interfacial DMI contribution $\mathbf{h}_{DMI}$ is obtained from the functional derivative of the DMI energy density $\varepsilon_{DMI} = D[m_z \nabla\cdot\mathbf{m} - (\mathbf{m}\cdot\nabla)m_z]$ under the hypothesis of thin film $\left(\frac{\partial \mathbf{m}}{\partial z} = 0\right)$ [1,9] as

$$\mathbf{h}_{DMI} = -\frac{2D}{\mu_0 M_S}\left[(\nabla\cdot\mathbf{m})\hat{z} - \nabla m_z\right], \quad (S9)$$

with $D$ being the parameter taking into account the intensity of the DMI, $m_z$ the out-of-plane component of the normalized magnetization, $\mu_0$ the vacuum permeability, and $\hat{z}$ the unit vector along the out-of-plane direction. The DMI affects the boundary conditions of the ferromagnetic sample in the following way $\frac{d\mathbf{m}}{dn} = \frac{D}{2A}(\hat{z}\times\mathbf{n})\times\mathbf{m}$, where $\mathbf{n}$ is the unit vector normal to the edge, and $A$ is the exchange constant.

The thermal effects are accounted in Eq. (S8) as a stochastic term $\mathbf{h}_{th}$ added to the deterministic effective magnetic field in each computational cell $\mathbf{h}_{th} = (\chi/M_S)\sqrt{2(\alpha K_B T / \mu_0 \gamma_0 \Delta V M_s \Delta t)}$, with $K_B$ being the Boltzmann constant, $\Delta V$ the volume of the computational cubic cell, $\Delta t$ the simulation time step, $T$ temperature of the sample, and $\chi$ a three-dimensional white Gaussian noise with zero mean and unit variance [10,11]. The noise is assumed to be uncorrelated for each computational cell. The discretization cell size used is 2.5x2.5x0.8 nm$^3$.

We nucleate a skyrmion with a negative out-of-plane core (region where $-1 \leq m_z < 0$) using the method proposed in Ref. [12] and calculate the area, perimeter and diameter $2R_d = D_d$ of the skyrmion in a post-process from saved snapshots of the magnetization texture.

Figs. S5 and S6 display histograms of the statistical distribution of the skyrmion diameter and perimeter, respectively, for different external magnetic fields and temperatures. All the histograms are characterized by a Gaussian distribution, but the application of the field or reduction of the temperature significantly decrease both mean and standard deviation, as indicated in the figures.

In order to micromagnetically obtain the scaling exponent $\gamma$ of the uniaxial perpendicular anisotropy, we considered $\gamma$ as free parameter. We performed athermal (deterministic) micromagnetic simulations, where the thermal effect was included through the scaling relations of $A$(T), $K_u$(T) and $D$(T) rather than including the stochastic field $\mathbf{h}_{th}$. As can be seen in Fig. S7(a), the best fitting has been achieved by considering $\gamma$=3.585. This $\gamma$ and the temperature dependence of the saturation magnetization are the input parameters of the analytical theory.

If we consider the scaling relations as calculated from atomistic simulations, we find results in agreement with micromagnetic simulations until the skyrmion is in the metastable region far from the boundary with the ground state region (see Fig. S7(b) for $T$<200 K).

with scaled values of the macroscopic parameters (a) $\alpha$=1.5, $\beta$=1.5, $\gamma$=3.585, and (b) $\alpha$=1.5, $\beta$=1.5, $\gamma$=3.03.

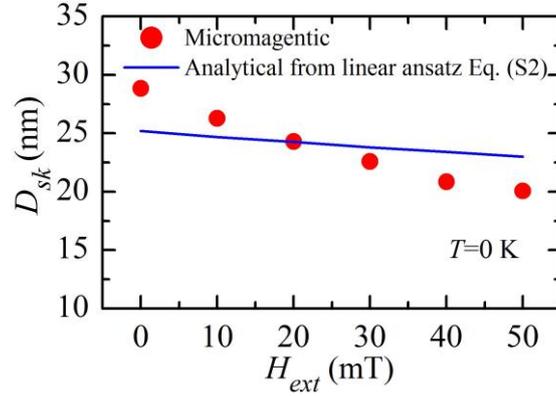

FIG. S1: Skyrmion diameter as a function of the perpendicular external field as computed by micromagnetic simulations (red circles) and by analytical computations based on the linear ansatz in Eq. (S2) (blue curve) at $T=0$ K.

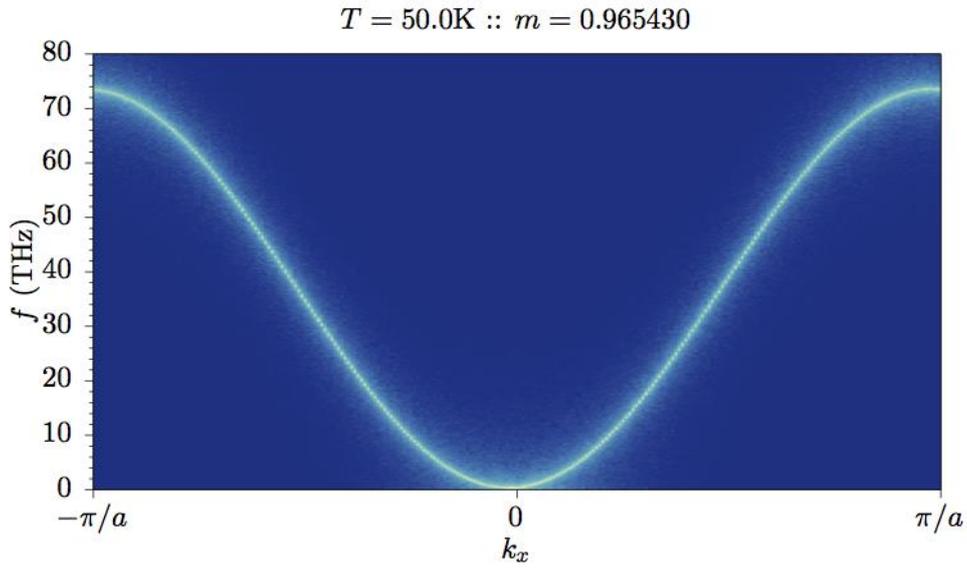

FIG. S2: Example of the spectrum calculated for a thin film using atomistic spin dynamics

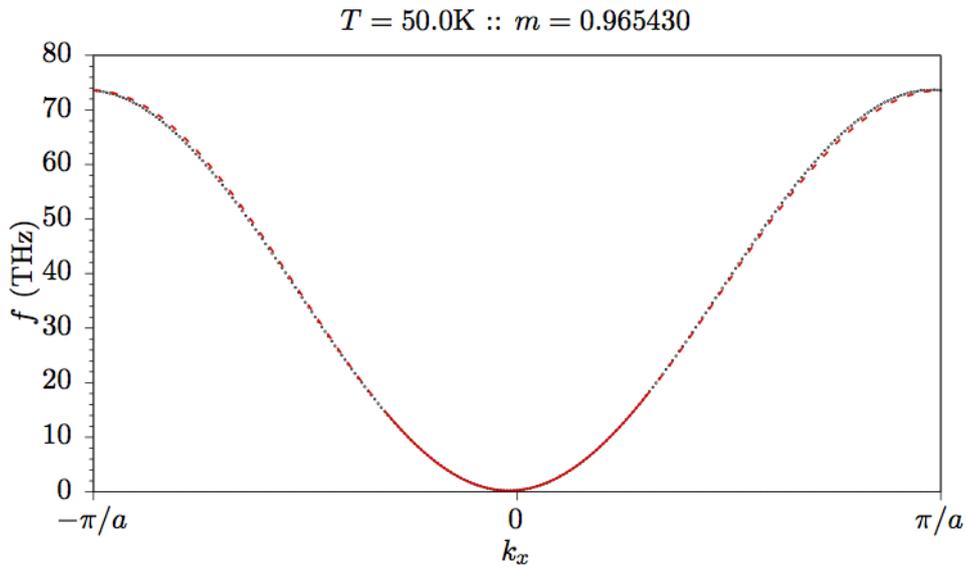

FIG: S3: Example of fitting of the linear spin wave dispersion relation. The dotted line is from the simulated data extracted by fitting a Lorentzian curve to the frequency data of each k-point. The red line is the fitted linear spin wave theory Eq. (6). The solid line indicates the long wavelength fitted region and the dash line is the continuation of the curve which is not fitted.

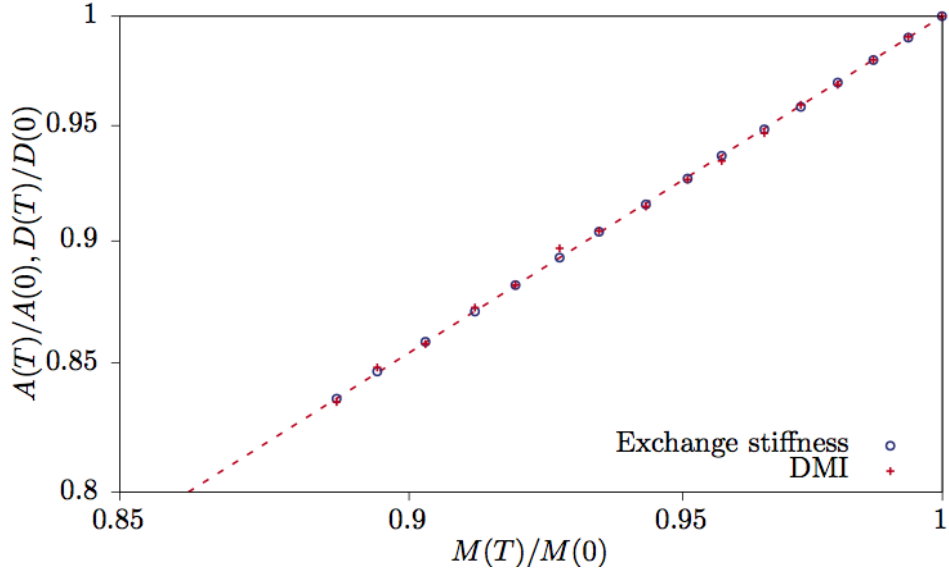

FIG: S4: Fitted scaling for the exchange stiffness $A$ ($\alpha$=1.498 +/- 0.001) and DMI strength $D$ ($\beta$=1.495 +/- 0.007)

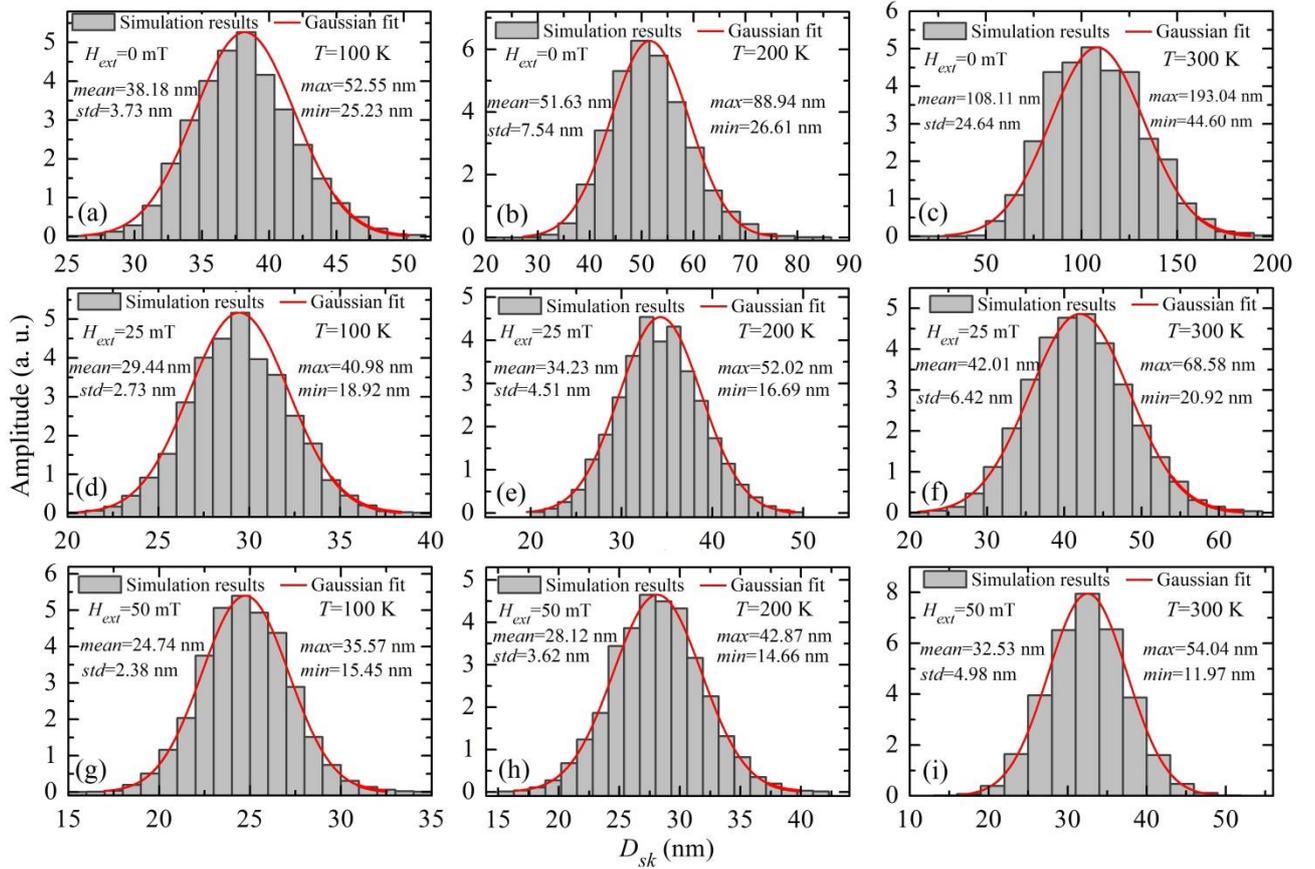

FIG: S5: Histogram representation of the statistical distribution of the skyrmion diameter and related Gaussian fit (red curve). In the rows, the external field is (a), (b), (c) $H_{ext}$=0 mT, (d), (e), (f) $H_{ext}$=25 mT, and (g), (h), (i) $H_{ext}$=50 mT. In the columns, the temperature is (a), (d), (g) *T*=100 K, (b), (e), (h) *T*=200 K, and (c), (f), (i) *T*=300 K. The notations mean, max, and min refer to the mean, maximum, and minimum value of the skyrmion diameter, while std is its standard deviation.

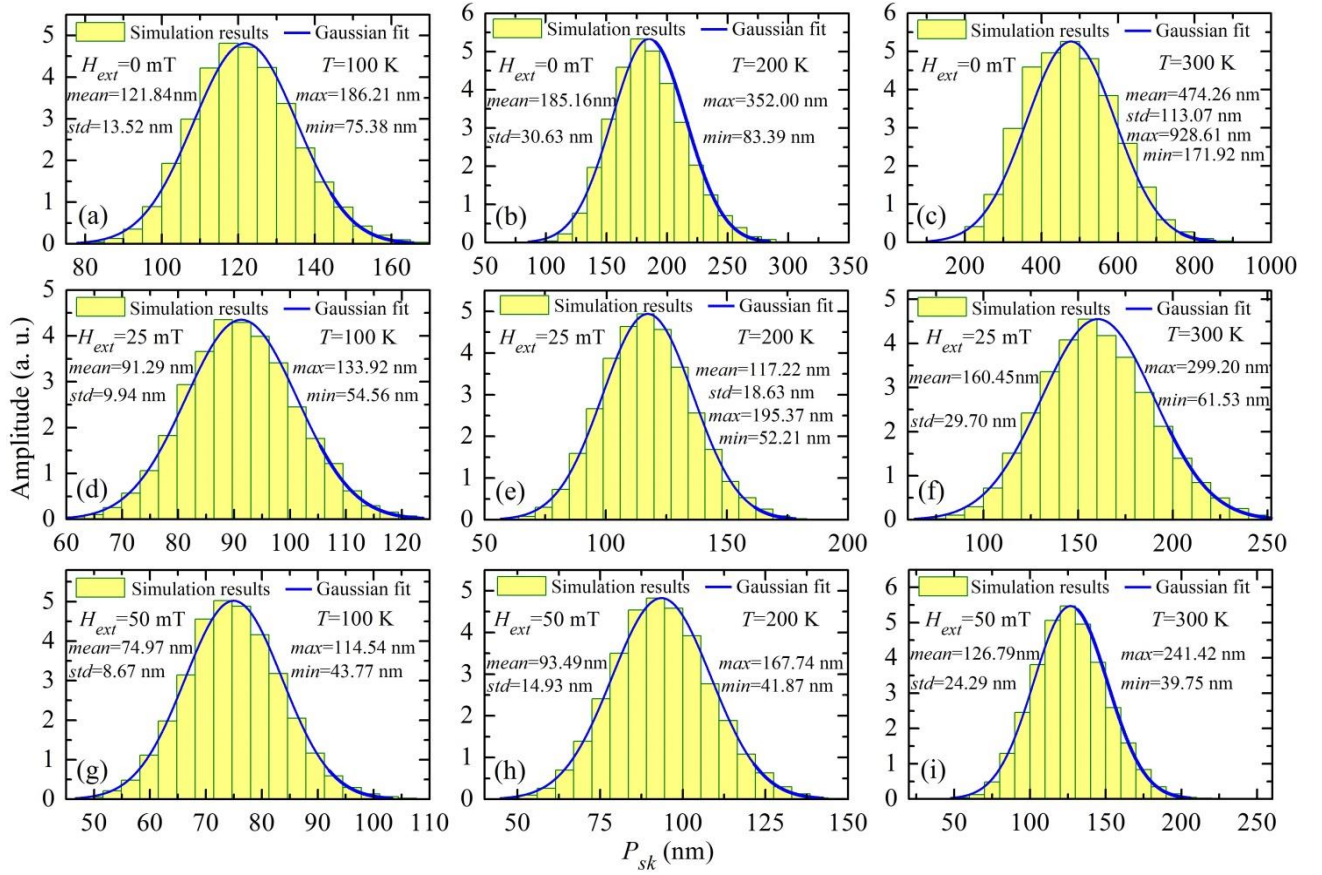

FIG. S6: Histogram representation of the statistical distribution of the skyrmion perimeter and related Gaussian fit (blue curve). In the rows, the external field is (a), (b), (c) $H_{ext}$=0 mT, (d), (e), (f) $H_{ext}$=25 mT, and (g), (h), (i) $H_{ext}$=50 mT. In the columns, the temperature is (a), (d), (g) *T*=100 K, (b), (e), (h) *T*=200 K, and (c), (f), (i) *T*=300 K. The notations mean, max, and min refer to the mean, maximum, and minimum value of the skyrmion perimeter, while std is its standard deviation.

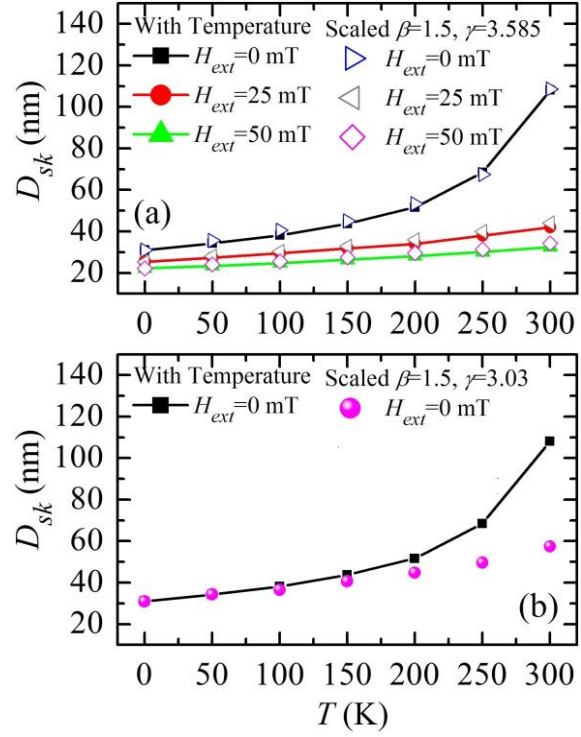

FIG. S7: Comparison between the mean value of the skyrmion diameter obtained by full micromagnetic simulations including thermal fluctuations (same as Fig. 1(b) in the main text) with the skyrmion diameter calculated from deterministic micromagnetic simulations